\documentclass[pra,twocolumn, amsmath,amssymb,aps,superscriptaddress,showpacs,nofootinbib]{revtex4-1}

\usepackage[utf8]{inputenc}	
\usepackage{graphicx}
\usepackage{epstopdf}
\usepackage{amsthm}
\usepackage{latexsym}
\usepackage{amsfonts}
\usepackage{color}

\usepackage{hyperref}
\usepackage{enumerate}
\usepackage{dcolumn}
\usepackage{bm}

\newcommand{\R}{\mathbb R} 

\newcommand{\C}{\mathbb C} 

\newcommand{\ip}[2]{\left\langle\,#1\,|\,#2\,\right\rangle} 
\newcommand{\tr}[1]{\textrm{tr}\left[#1\right]} 
\newcommand{\id}{\bm{1}} 
\newcommand{\mc}[1]{\mathcal{#1}} 

\newcommand{\ket}[1]{|{#1}\rangle}

\newcommand{\bra}[1]{\langle{#1}|}

\newcommand{\wvp}[3]{\,_{#3}\langle{#1}\rangle_{#2}^{w}}

\newcommand{\Ao}{\mathsf{A}}
\newcommand{\Po}{\mathsf{P}}
\newcommand{\Qo}{\mathsf{Q}}

\begin{document}
\title{Quantification of Concurrence via Weak Measurement}
\author{Mikko Tukiainen}
 \email{mikko.tukiainen@utu.fi}
\affiliation{%
 Turku Centre for Quantum Physics, Department of Physics and Astronomy, University of Turku, FI-20014, Turun Yliopisto, Finland
}%
\affiliation{%
Research Center of Integrative Molecular Systems (CIMoS), Institute for Molecular Science, National Institutes of Natural Sciences, Okazaki, Aichi 444-8585, Japan}%

\author{Hirokazu Kobayashi}
\email{kobayashi.hirokazu@kochi-tech.ac.jp}
\affiliation{%
Department of Electronic and Photonic System Engineering, Kochi University of Technology, Tosayamada-cho, Kochi 782-8502, Japan}%

\author{Yutaka Shikano}
\email{yshikano@qc.rcast.u-tokyo.ac.jp}
\affiliation{%
Research Center of Integrative Molecular Systems (CIMoS), Institute for Molecular Science, National Institutes of Natural Sciences, Okazaki, Aichi 444-8585, Japan}%
\affiliation{%
Institute for Quantum Studies, Chapman University, Orange, CA 92866, USA}%
\affiliation{%
Research Center for Advanced Science and Technology (RCAST), The University of Tokyo, Meguro, Tokyo 153-8904, Japan}%

\begin{abstract}
Since entanglement is not an observable {\it per se}, measuring its value in practice is a difficult task. 
Here we propose a protocol for quantifying a particular entanglement measure, namely concurrence, 
of an arbitrary two-qubit pure state via a single fixed measurement set-up by exploiting so-called weak measurements and the associated weak values together with the properties of the Laguerre-Gaussian modes. 
The virtue of our technique is that it is generally applicable for all two-qubit systems and does not involve simultaneous copies of the entangled state. 
We also propose an explicit optical implementation of the protocol.
\end{abstract}
\maketitle

\section{Introduction}
In the course of the past decades, the role of entanglement has evolved  
into a genuine quantum resource utilized in various quantum communication and computation protocols \cite{Bennett92, Bennett93, Ekert91, briegel2001, Horodecki09}. 
This evolution has been supported by the formidable progress made on the techniques of generating entanglement in practice.
Inevitable and inescapable noise, together with imperfections present in every real experiment, may, however, degrade the intended entangled state. 
Being able to detect and measure the entanglement content becomes important since any amount of entanglement can be harnessed in non-classical tasks \cite{Masanes2006, Masanes2008}.
Although several theoretical measures have been developed for this purpose \cite{Guhne09,Horodecki09}, realizing them in practice remains challenging in general. 
The reason is that typically these measures of entanglement contain rather involved, even unphysical, operations or are non-linear functions of the state. 

One of the most widely used measures of entanglement is the so-called concurrence \cite{Wootters98}, which in the case of two qubits in a pure state takes a particularly simple form. Despite the mathematical simplicity, the task of quantifying the value of concurrence of an {\it unknown} two-qubit pure state using only a single measurement set-up of a fixed normalized projection-valued measure (PVM) is impossible \cite{Sancho2000}.
Nevertheless, several different procedures circumventing this impossibility have been reported that exploit collective measurements done with simultaneous copies of the state \cite{Horodecki03, Mintert05, Walborn06, Zhang13} or utilize the curious relation between concurrence and two-particle 
interference \cite{Lee08}. Furthermore, measurements of concurrence that rely on relaxing the aforementioned PVM-criterion have been developed \cite{Rehacek04, Salles06, Sheng15}.

In this study, we propose a local tomographic strategy to quantify the concurrence of {\it any} two-qubit pure state 
that takes advantage of so-called weak measurements. We also consider an experimental implementation on an optical set-up 
that can be deployed to measure the concurrence of two polarization entangled photons using the proposed protocol. 
Our method is, however, universal in the sense that it works for all two-qubit systems.

The key tools of our proposal are weak measurements and the resulting weak values \cite{Aharonov87, Aharonov88}. Weak measurements are (von Neumann) standard measurements \cite{QTM96} 
where the coupling strength $\lambda$ between the measured system and the measuring pointer is minuscule. Consequently, the disturbance of the weak measurement to 
any subsequent (strong) measurement, usually called post-selection, is negligible. By post-selecting on a particular pure state $|\varphi\rangle\langle\varphi|$, in the vanishing interaction strength 
limit $\lambda \rightarrow 0$, one can derive the weak value of the observable $\Ao$ as
\begin{eqnarray}
_{\langle \varphi |} \langle \Ao \rangle_{\rho}^w := \frac{\tr{\Ao \, \rho\,  |\varphi \rangle\langle \varphi|}}{\tr{\rho\, |\varphi \rangle\langle \varphi|}},
\end{eqnarray}
where $\rho$ is the pre-selected (mixed) state of the measured system \cite{Wiseman02}. Throughout this paper, we omit the pre-selection sub-index whenever it is clear from the context. Weak values are intrinsically complex which has already proved useful in characterizing the mathematically observable-independent probability space~\cite{AS}, 
several quantum paradoxes~\cite{AR}, the quantum state~\cite{Lundeen11, Haapasalo11, Lundeen12, Wu13, Kobayashi14, Malik14, Leach}, and 
unobservable quantities such as the geometric phase~\cite{Sjoqvist06, SA, Kobayashi10, Kobayashi11} 
and the non-Hermitian operator \cite{Pati15}; see also the review papers \cite{AV08, Shikano_rev, Dressel_rev, Nori_rev}. We show that one may also take advantage of the complex feature of the weak values in assessing the amount of entanglement with a single measurement set-up.
This result builds up on the fact first noted in Ref.\,\cite{Kobayashi14} that, when a Laguerre-Gaussian beam is used as the pointer state of the weak measurement, certain weak values can 
be interpreted as stereographical projections of the Bloch sphere onto $\R^2$-plane.

\section{Concurrence and Weak Values}
Let us assume that two observers, Alice and Bob, are tasked with determining the amount of entanglement in a bipartite state $\rho_{AB}$ by means of performing local operations. 
Furthermore, assume that the source generates only pure two-qubit states, that is, $\rho_{AB}=|\Psi_{AB}\rangle\langle \Psi_{AB}|$ for some
\begin{eqnarray}
|\Psi_{AB}\rangle = a_{00} |00\rangle + a_{01} |01\rangle + a_{10} |10\rangle + a_{11} |11\rangle,
\end{eqnarray}
where $|0\rangle$ and $|1\rangle$ are the eigenvectors of Pauli operator $\sigma_z$, and $|ij\rangle := |i\rangle \otimes |j\rangle$ and $a_{ij}$ (${i,j=0,1}$) are complex numbers satisfying the normalization $\sum_{i,j=0}^1 |a_{ij}|^2 = 1$. One of the most widely used entanglement measures in two-qubit systems is the {\it concurrence} $C$. In the case of a pure state $|\Psi_{AB}\rangle$, the concurrence $C(\Psi_{AB})$ takes the simple form \cite{Wootters98}
\begin{eqnarray}
C(\Psi_{AB})^2 &=& 4 |a_{00}a_{11}-a_{01}a_{10}|^2 \nonumber \\
&=& 4\det(\rho_A) = 4 \det(\rho_B),
\end{eqnarray} where $\rho_{A(B)}$ is the reduced density matrix of Alice (Bob), {\it e.g.,}
\begin{eqnarray}
\rho_A &=& \left( \begin{array}{cc}
|a_{00}|^2 + |a_{01}|^2 & a_{00}^* a_{10} + a_{01}^* a_{11} \\
a_{00} a_{10}^* + a_{01} a_{11}^* & |a_{10}|^2 + |a_{11}|^2
\end{array} \right).
\end{eqnarray}
The concurrence has a one-to-one connection to the von Neumann entropy \cite{Wootters98}
\begin{eqnarray}\label{vneumann}
E(\Psi_{AB})&=&-\tr{\rho_A \log_2(\rho_A)}=-\tr{\rho_B \log_2(\rho_B)}\nonumber \\
&=&-\frac{1+\sqrt{1-C^2}}{2}\log_2\left(\frac{1+\sqrt{1-C^2}}{2}\right) \nonumber \\
& &- \frac{1-\sqrt{1-C^2}}{2}\log_2\left(\frac{1-\sqrt{1-C^2}}{2}\right),
\end{eqnarray} 
and via that to a plethora of other entanglement measures \cite{Horodecki09,Guhne09}, which makes it a natural choice of figure of merit for our task.

\begin{figure*}[t!]
\begin{minipage}[t]{0.4\textwidth}
\begin{flushleft}(a)\end{flushleft}
\vspace*{-0.4cm}
\includegraphics[width=0.95\textwidth]{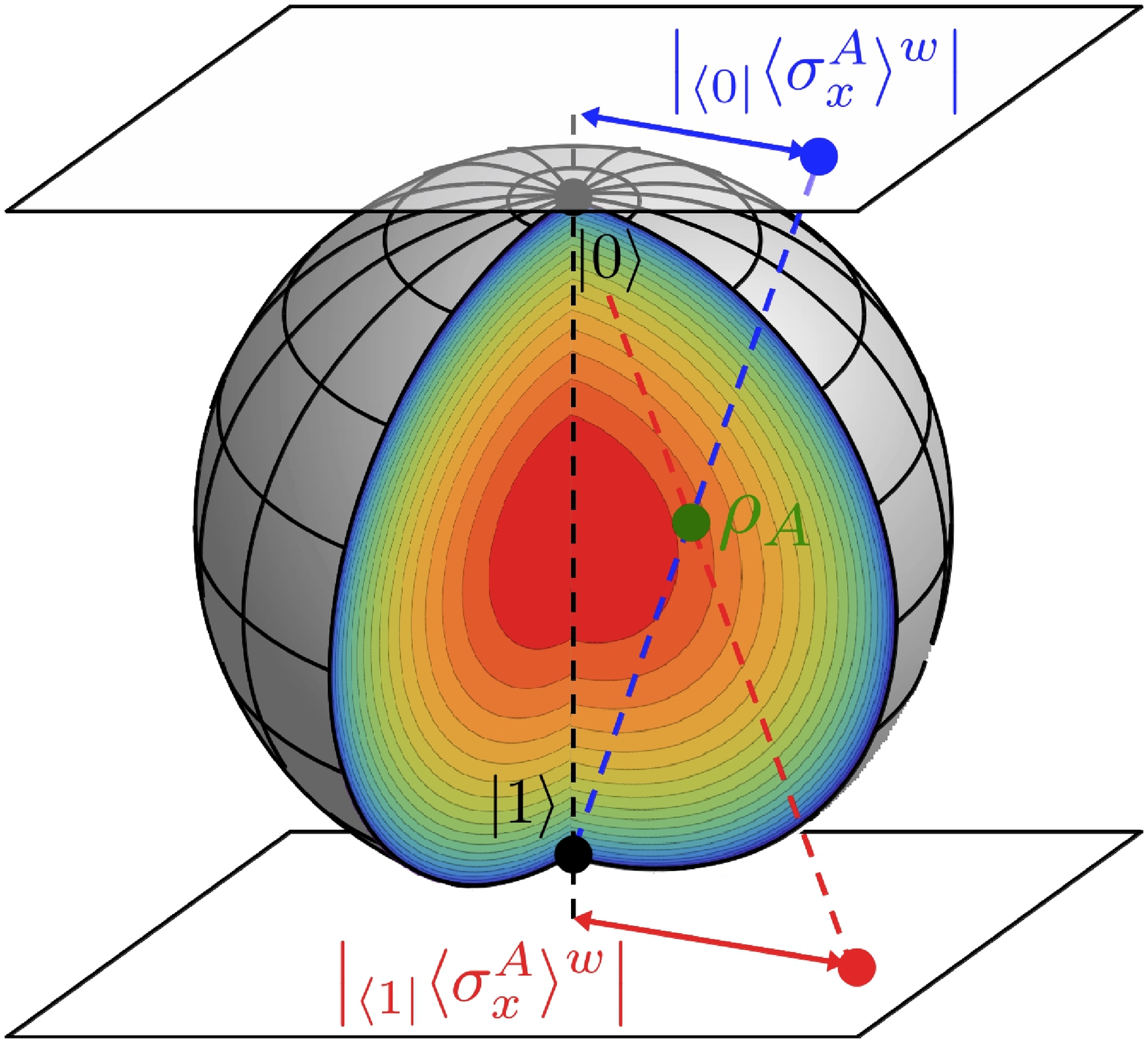} 
\end{minipage}
\begin{minipage}[t]{0.55\textwidth}
\begin{flushleft}(b)\end{flushleft}
\vspace*{-1.2cm}
\includegraphics[width=0.9\textwidth]{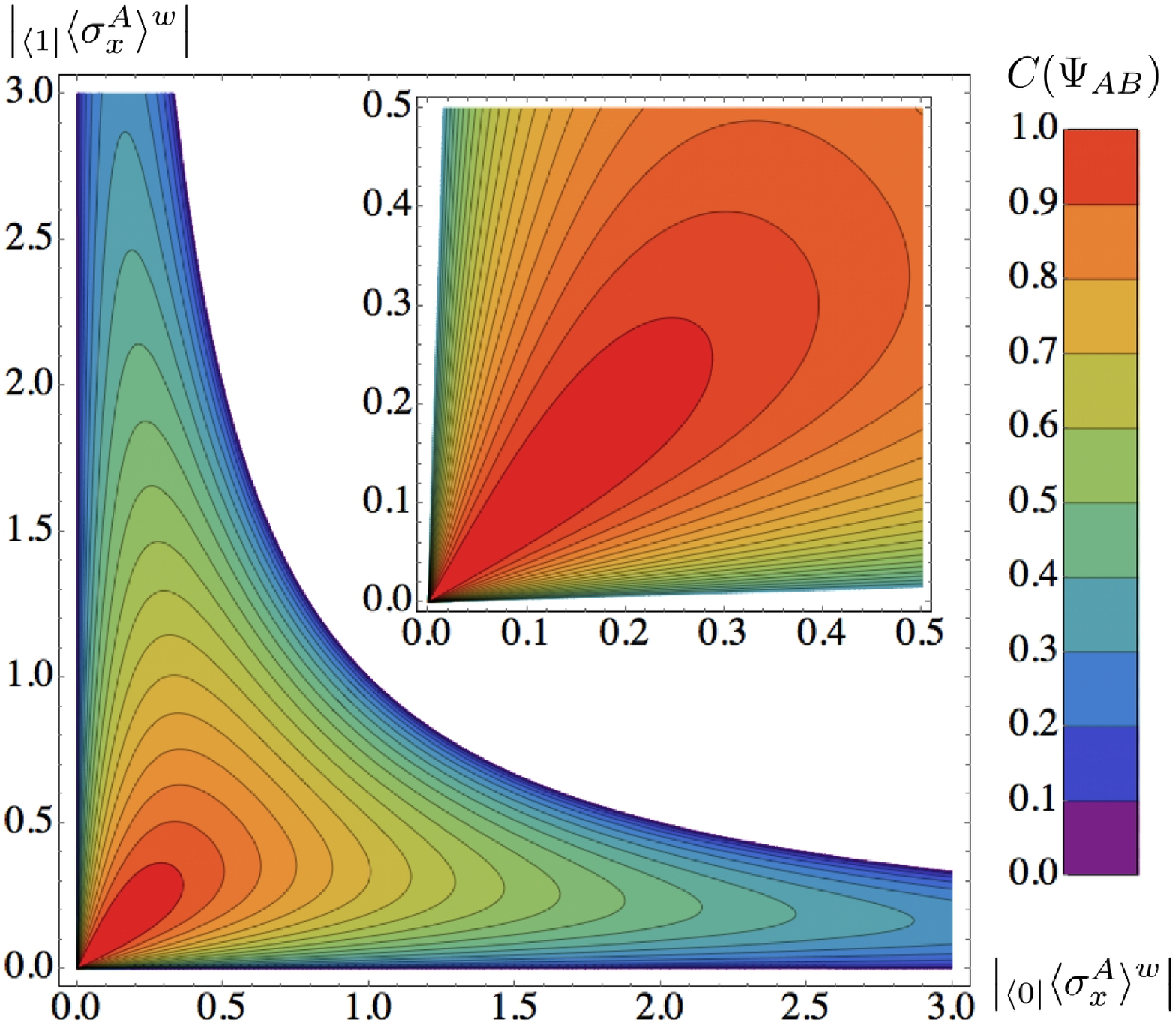}
\end{minipage}
\caption{ (a) Stereographical representation of the weak value of the state $\rho_A$. Following Ref.\,\cite{Kobayashi14}, the weak measurement of $\sigma_x$ on the state $\rho_A$, followed by post-selection 
on $|0\rangle$ or $|1\rangle$, may be interpreted as stereographic projections of the qubit state $\rho_A$ on the two $\R^2$-planes that intersect the north and south pole of the Bloch sphere. 
For our purposes, the absolute values $|\wvp{\sigma_x^A}{}{\bra{0}}|$ and $|\wvp{\sigma_x^A}{}{\bra{1}}|$ are particularly important because they may be used to measure the distance between $\rho_A$ 
and the maximally mixed state $\frac{1}{2} \id_A$, which in turn is related to the amount of entanglement. 
(b) Concurrence $C(\Psi_{AB})$ in terms of $|\wvp{\sigma_x^A}{}{\bra{0}} |$ and $|\wvp{\sigma_x^A}{}{\bra{1}}|$. The concurrence is fully determined by these variables except for the point $(0,0)$, which 
corresponds to the black dashed line in (a). The white region equals to the canceled area $|\wvp{\sigma_x^A}{}{\bra{0}}| |\wvp{\sigma_x^A}{}{\bra{1}}| > 1$.}
\label{fig:weak}
\end{figure*}

Our main result is to reveal a mathematical relationship between the concurrence and the weak values corresponding to weak measurements of either one of the local observers. For instance, Alice's weak values of the observable $\sigma_x^A := |0\rangle\langle 1|+|1\rangle\langle 0|$, pre-selected on her reduced state $\rho_A$ and post-selected on either $|0\rangle$ or $|1\rangle$, read  
\begin{eqnarray}\label{eq:wvs}
\,_{\langle 0 |} \langle \sigma_x^A \rangle^w &:=& \frac{\tr{\sigma^A_x\, \rho_A   \, |0\rangle\langle 0|}}{\tr{\rho_A \, |0\rangle\langle 0|}}= \frac{a_{00} a_{10}^* + a_{01} a_{11}^*} {|a_{00}|^2+|a_{01}|^2},  \nonumber \\
\wvp{\sigma_x^A}{}{\bra{1}} &:=& \frac{\tr{ \sigma_x^A \, \rho_A \,|1\rangle\langle 1|}}{\tr{\rho_A \, |1\rangle\langle 1|}} = \frac{a_{00}^* a_{10} + a_{01}^* a_{11}} {|a_{10}|^2+|a_{11}|^2};
\end{eqnarray} 
see Fig.\,\ref{fig:weak}(a).
A weak value may not be well-defined, if its denominator vanishes. Physically this corresponds to
receiving no signal on the measuring pointer whatsoever. 
We notice that either one of the above weak values being non-vanishing automatically implies that the other one is also non-zero. 
Therefore, whenever ${\wvp{\sigma_x^A}{}{\bra{0}}\neq 0 \neq \wvp{\sigma_x^A}{}{\bra{1}}}$, we may write ${\left(|a_{00}|^2+|a_{01}|^2\right)/\left(|a_{10}|^2+|a_{11}|^2\right) = |\wvp{\sigma_x^A}{}{\bra{0}}| / | \wvp{\sigma_x^A}{}{\bra{1}}|}$ and solve
\begin{widetext}
\begin{eqnarray}\label{eq:concwv}
C(\Psi_{AB})^2 = 4 \det( \rho_A) 
= 4\left(1- |\wvp{\sigma_x^A}{}{\bra{0}}| |\wvp{\sigma_x^A}{}{\bra{1}}| \right) \frac{|\wvp{\sigma_x^A}{}{\bra{0}}| |\wvp{\sigma_x^A}{}{\bra{1}}|}{\left(|\wvp{\sigma_x^A}{}{\bra{0}}|+|\wvp{\sigma_x^A}{}{\bra{1}}|\right)^2},
\end{eqnarray}
\end{widetext}
where we have used the information $\wvp{\sigma_x^A}{}{\bra{0}} \wvp{\sigma_x^A}{}{\bra{1}} = |\wvp{\sigma_x^A}{}{\bra{0}}| |\wvp{\sigma_x^A}{}{\bra{1}}|$. 
Since $ |a_{00} a_{10}^* + a_{01} a_{11}^*| \leq |a_{00}| |a_{10}| + |a_{01}| |a_{11}| \leq  1/2$, 
we additionally conclude that 
\begin{eqnarray}
 |\wvp{\sigma_x^A}{}{\bra{0}}| |\wvp{\sigma_x^A}{}{\bra{1}}| \leq 1.
\end{eqnarray}
On the other hand, one of the weak values in Eq.\,(\ref{eq:wvs}) being zero implies that the other one either also vanishes or is not well-defined. Assume for example that $\wvp{\sigma_x^A}{}{\bra{0}}$ is not well-defined. 
Then $|a_{00}|^2 +|a_{01}|^2 = 0$ implying $C(\Psi_{AB})=0$. Similarly, $C(\Psi_{AB})=0$ if $\wvp{\sigma_x^A}{}{\bra{1}}$ is not well-defined. These observations can also be reproduced from 
Eq.\,(\ref{eq:concwv}) as limiting cases $\wvp{\sigma_x^A}{}{\bra{i}}\to \infty$, $i=0,1$, since, except for the point $\left(|\wvp{\sigma_x^A}{}{\bra{0}} |, |\wvp{\sigma_x^A}{}{\bra{1}}|\right)=(0,0)$, concurrence $C(\Psi_{AB})$ is a continuous function of $|\wvp{\sigma_x^A}{}{\bra{0}}|$ and $|\wvp{\sigma_x^A}{}{\bra{1}}|$ 
\footnote{Actually, in the point $(0,0)$ concurrence is not even a function of two weak values. Namely, with a proper choice 
of $\Psi_{AB}$ $C(\Psi_{AB})$ can acquire any value from the interval $[0,1]$, while satisfying ${\wvp{\sigma_x^A}{}{\bra{0}}= 0 = \wvp{\sigma_x^A}{}{\bra{1}} }$.}. 
Therefore, the concurrence, plotted in 
Fig.\,\ref{fig:weak}(b), may be determined from Eq.\,(\ref{eq:concwv}), with the exception of the singularity in the origin $\left(|\wvp{\sigma_x^A}{}{\bra{0}} |, |\wvp{\sigma_x^A}{}{\bra{1}}|\right)=(0,0)$; this case will be later analysed separately. It is noteworthy that the protocol presented works completely locally.


We note in passing that $|\wvp{\sigma_x^A}{}{\bra{0}} | = |\wvp{\sigma_x^A}{}{\bra{1}}|$ is the only line passing through the origin on which $C(\Psi_{AB})$ attains its maximum value $1$. The reduced states $\rho_A$ corresponding to this line are those which are on the equatorial plane of the Bloch sphere in Fig.\,\ref{fig:weak}(a). On this line, Eq.\,(\ref{eq:concwv}) simplifies to 
\begin{eqnarray}
C(\Psi_{AB}) = \sqrt{1- | \wvp{\sigma_x^A}{}{\bra{0}} |^2}.
\end{eqnarray}
This observation is useful in order to calibrate $C$ as close to unity (or any other value from the interval $[0,1]$) as desired. 
Because the process is completely local, the other party (Bob) can validate the result of this ``optimization'' for instance via state tomography. 

\section{Determination of entanglement with a fixed measurement set-up}
Determining $C(\Psi_{AB})$ of arbitrary $|\Psi_{AB}\rangle$ directly via measurement of only a single set of orthogonal projectors $\Po_i = |O_i\rangle\langle O_i|$, $\sum_{i=1}^4 \Po_i =\id$, 
where $\langle O_i | O_j \rangle = \delta_{ij}$ (Kronecker delta), is impossible \cite{Sancho2000}. In other words, one cannot quantify concurrence of all bipartite states with a single fixed measurement set-up if the measured observable is a PVM. This is due to the fact that the measured probabilities $p_i = |\ip{O_i}{\Psi_{AB}}|^2$ result in three independent real numbers, which are not in general sufficient to determine $C(\Psi_{AB})$, a non-linear function of four complex parameters. In fact, even deciding if a completely unknown (hence possibly mixed) bipartite state is the entanglement or not requires as many resources as state tomography \cite{Carmeli15}.

The relationship between Alice's weak values and the concurrence introduced in the previous section suggests that weak measurements allow one to circumvent this impossibility. 
To extract the real and imaginary parts of the weak value 
two complementary pointer observables are usually used \cite{Aharonov88, Aharonov90, Jozsa, Resch04}, that is, two separate measurements have to be set up. Remarkably however, it is also possible to quantify both of these components simultaneously by using so-called Laguerre-Gaussian (LG) modes \cite{Puentes12, Kobayashi12, Kobayashi14} as the initial pointer state due to the initial correlations \cite{Home16} related to these states.

As alluded in the previous section, the determination of entanglement fails only in problematic
cases where $\wvp{\sigma_x^A}{}{\bra{0}}= 0 = \wvp{\sigma_x^A}{}{\bra{1}}$. The vanishing weak values imply that Alice's state is simplified to
\begin{eqnarray}\label{eq:problemstate}
\rho_A &=& \left( \begin{array}{cc}
|a_{00}|^2 + |a_{01}|^2 &0 \\
0 & |a_{10}|^2 + |a_{11}|^2
\end{array} \right).
\end{eqnarray}
These cases correspond to the states on a line connecting the opposite poles $|0\rangle$ and $|1\rangle$ of the Bloch sphere [see Fig.\,\ref{fig:weak}(a)]. In our protocol the set of these states has only minor relevance since mathematically it has null measure (in the relevant measurable space). Accordingly, the impossibility of determining the concurrence of states with a single PVM strategy persists even if these problematic states are excluded. 
Nevertheless, in such instances a local measurement of the 
post-selection probabilities can be used to reveal the amount of entanglement in the state $|\Psi_{AB}\rangle$. To this end, Alice can measure the relative intensities of the post-selected states to solve the diagonal elements of 
$\rho_A$ in Eq.\,(\ref{eq:problemstate}). Since this measurement may be done jointly with the weak measurement protocol described above, the whole procedure of determining the entanglement content in $|\Psi_{AB}\rangle$ can be achieved with a single fixed measurement device. Moreover, the protocol uses only a single fixed PVM as the post-selected measurement; as discussed above, without the preceding weak interaction such an entanglement-measuring strategy would be impossible. 

The weak values $\wvp{\sigma_x^A}{}{\bra{0}}$ and $\wvp{\sigma_x^A}{}{\bra{1}}$, 
in addition to the intensity measurements described above, give sufficient information to determine the reduced state $\rho_A$ [see Fig.\,\ref{fig:weak}(a)]. In this regard, the protocol we presented essentially relies on local tomography of the reduced state of Alice (or Bob): in the absence of classical communication between the two parties, as is the case in our protocol, this is the optimal local strategy to determine the entanglement of the two-qubit state \cite{Sancho2000}. As a consequence, we have generalized the one-qubit pure state tomography described in Ref.\,\cite{Kobayashi14} for mixed states thus expanding the scope of applications of the previously introduced technique. This underpins that the weak measurement set-up exploiting Laguerre-Gaussian modes (an optical proposal of which is given in the next section) could be considered as a basic tool in quantum experiments involving tomography of qubits, such as the above introduced entanglement quantification. 

Being able to reconstruct the reduced state $\rho_A$ will also enable one to connect the weak values to entanglement measures other than concurrence; see for example Eq.\,\eqref{vneumann}. In fact, the von Neumann entropy $E(\Psi_{AB}) = -\tr{ \rho_{A(B)} \log_2(\rho_{A(B)})}$ in Eq.\,\eqref{vneumann} is an entanglement measure of, not only two-qubits, but also general bipartite pure states $\Psi_{AB}$ and independent of which one of the subsystems $A$ or $B$ it is calculated with respect to. Hence, our protocol can be immediately generalized to assess the amount of entanglement in scenarios, where one of the two parties possesses an one-qubit system. Additionally, our method is not fully confined to pure states but can also be utilised in estimating the entanglement of the mixed bipartite states; we have left the details and proof of this fact to Appendix \ref{appen}. This is a highly important upside from the experimentalists' viewpoint, since the preparation of a perfectly pure bipartite state is not a realistic assumption in practice.

\section{Proposal for optical experiment}\label{sec:proposal}
\def\bra#1{\langle #1|}
\def\ket#1{| #1\rangle}
\def\bracket#1{\langle\mbox{$#1$}\rangle}
\def\bracketi#1#2{\langle\mbox{$#1$}|\mbox{$#2$}\rangle}
\def\bracketii#1#2#3{\langle\mbox{$#1$}|\mbox{$#2$}|\mbox{$#3$}\rangle}
\def\sub#1{_{\scriptsize\mbox{#1}}}
\def\sur#1{^{\scriptsize\mbox{#1}}}
In this section we describe a possible optical set-up for 
determining the concurrence of the polarization entangled state
$|\Psi_{AB}\rangle$ of photon pairs via a weak
measurement. Our proposed experimental set-up is illustrated in Fig.\,\ref{fig:setup}. 
The reduced density matrix $\rho_A$ weakly interacts
with the pointer state via the interaction 
\begin{eqnarray}
U_\lambda = e^{-i \lambda \sigma_x \otimes \Po_x} = \Pi_+ \otimes e^{-i \lambda \Po_x}+\Pi_- \otimes e^{ i \lambda \Po_x},
\label{eq:interaction}
\end{eqnarray} 
where $\Po_x$ is the momentum operator along the $x$-direction on the 
cross-sectional plane of the optical beam, $\lambda$ is
a small interaction strength, and $\Pi_\pm = \frac{1}{2}(\id \pm
\sigma_x)$ are the eigenprojectors of the Pauli operator $\sigma_x$. 
The interaction (\ref{eq:interaction}) can be implemented using a polarization Sagnac
interferometer and the interaction strength $\lambda$ can be changed by tilting the
angle of a mirror inside the interferometer [see the inset in Fig.~\ref{fig:setup}]. 
\begin{figure}[h!]
\includegraphics[width=0.45\textwidth]{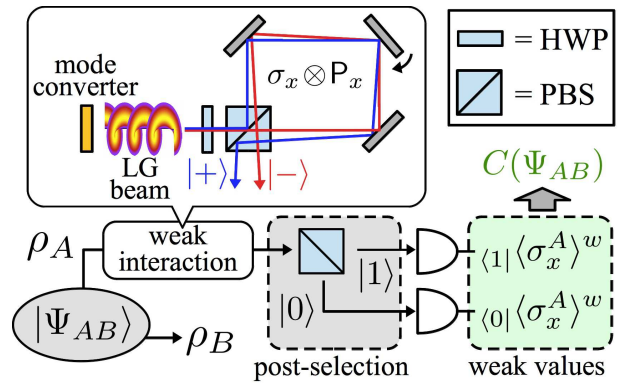}
\caption{Weak measurement set-up for determining concurrence in the two-photon polarization state $|\Psi_{AB}\rangle$. The initial pointer state is prepared as the Laguerre-Gaussian (LG) mode using a mode converter. The
 weak interaction between eigenvectors of $\sigma_x$,
 $\ket{\pm}=(\ket{0}\pm\ket{1})/\sqrt{2}$, 
can be implemented using a polarization Sagnac interferometer (PBS: polarization beam splitter, HWP: half waveplate).}
\label{fig:setup}
\end{figure}

As the initial pointer state, we choose the optical propagation mode with two-dimensional
normalized amplitude distribution $\phi\sub{i}(x,y)$, which satisfies the paraxial
wave equation \cite{siegman1986lasers}. 
After weak interaction and post-selection onto $\ket{\varphi}$ $(=\ket{0}$ or $\ket{1})$,
the intensity distribution $I\sub{f}(x,y)$ of the final pointer state becomes
\begin{eqnarray}\label{eq:intensity}
&& I^\varphi\sub{f}(x,y)\nonumber\\
=&&\sum_{j,k=\pm 1}\bracketii{\varphi}{\Pi_j\rho_A\Pi_k}{\varphi}\phi\sub{i}(x-j\lambda,y)\phi\sub{i}^*(x-k\lambda,y).
\end{eqnarray}
Assuming the ``weakness'' condition, ${\lambda^{-1} \gg \max\left( 1,
|\wvp{\sigma_x}{}{\bra{\varphi}}|\right)}$, the interaction in Eq.~(\ref{eq:interaction}) induces a translational shift of the pointer state with an amount proportional 
to the weak value $\wvp{\sigma_x}{}{\bra{\varphi}}$ along the $x$ direction \cite{Aharonov90}. 
Namely, under the weakness condition, Eq.~(\ref{eq:intensity}) can be approximated as
\begin{eqnarray}
&&I^\varphi\sub{f}(x,y)
=I^\varphi\sub{tot} \left|\phi\sub{i}(x-\lambda\wvp{\sigma_x}{}{\bra{\varphi}},y)\right|^2
\label{eq:wm-probe},
\end{eqnarray}
where $I^\varphi\sub{tot} \equiv \int dx \, dy \, I^\varphi\sub{f}(x,y) = \bra{\varphi}\rho_A \ket{\varphi}$ corresponds to the total intensity of the post-selected beams.

If the fundamental Gaussian beam is used for the pointer state, we can
extract only the real part of the weak value from the shift in the beam average position
and an alternative measurement set-up with additional optical components is required to
obtain the imaginary part of the weak value from the shift in the beam average momentum. A more suitable choice for the pointer
state for our purpose is the (first order) LG beam
$\phi\sub{i}(x,y)\propto (x+iy)\exp[-(x^2+y^2)]$, which is
a cylindrically symmetric solution of the paraxial wave equation
\cite{siegman1986lasers,allen1992orbital}. The LG beam can be generated from a Gaussian one by using a mode converter, such as a q-plate \cite{marrucci2006optica} or a spatial light modulator \cite{ando2009mode}. From Eq.\,(\ref{eq:wm-probe}), the averaged value of the position operators
$\Qo_x$ and $\Qo_y$ on the cross-sectional plane of the final
intensity distribution are calculated as
\begin{eqnarray}
\langle \Qo_x \rangle_f = \lambda \, \text{Re}[\wvp{\sigma_x}{}{\bra{\varphi}}]
\,,\,
\langle \Qo_y \rangle_f &&= \lambda \, \text{Im}[\wvp{\sigma_x}{}{\bra{\varphi}}]. \quad
\end{eqnarray}
Using a two-dimensional image sensor as a detector the LG pointer state therefore 
allows us to simultaneously visualize both the real and the imaginary part of the weak values $\wvp{\sigma_x}{}{\bra{0}}$ and
$\wvp{\sigma_x}{}{\bra{1}}$ without additional optical components \cite{Kobayashi14}. 

In the case of vanishing weak values, where Eq.\,(\ref{eq:concwv}) cannot be used, 
we cannot obtain any information about the entanglement from the averaged shifts of the pointer state. 
However, due to the aforementioned reasons these cases are physically insignificant. For the sake of completeness we nevertheless point out that measuring the total intensities $I^\varphi\sub{tot}=\bra{\varphi} \rho_A \ket{\varphi}$ of the two post-selected beams with $\ket{\varphi}=|0\rangle,|1\rangle$ enables 
one to determine the diagonal elements of the state in Eq.\,(\ref{eq:problemstate}). Because this can be performed jointly with measurements of the $\Qo_x$ and $\Qo_y$ position operators, 
one can determine the concurrence of the quantum state $|\Psi_{AB}\rangle$ with a single measurement set-up.

Although the LG mode pointer states allow us to determine the concurrence using a single fixed PVM for post-selection, there are some technical difficulties. 
The first problem is the mode conversion from the fundamental Gaussian mode to the LG mode. 
The conversion efficiency is limited by the mode converter and also by the mode coupling coefficient
between the incident mode of the photon pairs and the LG mode. 
To increase the mode-coupling coefficient, a single mode optical fiber is
typically used for spatial mode cleaning of the photon pair beam. 
In this case, however, fiber coupling loss becomes a serious problem for photon-pair detection.
One practical solution is photon-pair generation via four-wave mixing in the single mode fiber \cite{li2005optical}. 
Another problem is the low detection efficiency of the typical image sensor and, concurrently, the demand for a large ensemble of states needed to extract the weak values. 
To obtain the high-contrast two-dimensional intensity distribution, 
we have to generate photon pairs with high intensity using pulsed light or
a high-gain imaging sensor, such as a cascade of single photon detectors. 

\section{Summary}
We have shown how weak measurements and weak values can be used to quantify the concurrence of any two-qubit pure state. We demonstrated that the proposed protocol can be performed 
with a single measurement set-up using a local weak interaction and a Laguerre-Gaussian mode as the pointer state. Notably, the protocol uses a single fixed PVM as for the post-selection. In contrast, without the preceding weak interaction, such a measurement of concurrence is impossible \cite{Sancho2000}. We also considered a potential experimental realization for quantifying the concurrence of the polarization entangled state of photon pairs. Although the proposed implementation has some technical difficulties, such as the detection efficiency, we believe that our protocol could be practically implemented and demonstrated in the future.

\section*{Acknowledgments}
M.\,T. acknowledges financial support from the University of Turku Graduate School (UTUGS) and 
the hospitality of the Institute for Molecular Science (IMS), National Institutes of Natural Sciences (NINS) under the 
financial support of the IMS internship program. 
Y.\,S. was supported in part by Perimeter Institute for Theoretical Physics. Research at Perimeter Institute is supported 
by the Government of Canada through Industry Canada by the Providence of Ontario through the Ministry of Economic Development \& Innovation.
This work was supported by a Grant from The Murata Science Foundation, The Matsuo Foundation, IMS Joint Study Program, 
NINS youth collaborative project, the Center for the Promotion of Integrated Sciences (CPIS) of Sokendai, 
ICRR Joint Research from The University of Tokyo, JST ERATO (Grant No. JPMJER1601), 
and JSPS KAKENHI Grant Numbers 25790068, 25287101, and 16K05492.


\appendix
\section{Robustness of concurrence} \label{appen}
Our protocol relies on the fact that for a pure state $|\Psi_{AB}\rangle \in \C^2\otimes \C^2$ the concurrence is related to the reduced state $\rho_A:=\text{tr}_B\left[|\Psi_{AB}\rangle \langle \Psi_{AB}| \right]$ via $C(\Psi_{AB})^2 = 4 \det(\rho_A)$; see Eq.\,\eqref{eq:concwv}. The concurrence of a mixed two-qubit state $\rho_{AB}$ can then be obtained from the convex roof extension $C(\rho_{AB}) = \inf_{\{p_i, \psi_i\}} \sum_i p_i C(\psi_i)$, where $p_i \geq 0$ satisfies $\rho_{AB} = \sum_i p_i |\psi_i\rangle\langle \psi_i|$ for some unit vectors $|\psi_i\rangle \in \C^2\otimes \C^2$ \cite{Horodecki09}. In general $C(\rho_{AB})^2\neq 4 \det(\zeta_A)$, where $\zeta_A = \text{tr}_{B}\left[\rho_{AB}\right]$. Nevertheless, in this Appendix we will show that $C(\rho_{AB})^2 \approx 4 \det(\zeta_A)$ is a good estimate, whenever $\rho_{AB} \approx |\Psi_{AB}\rangle \langle \Psi_{AB}|$ for some unit vector $|\Psi_{AB}\rangle \in \C^2\otimes \C^2$. More precisely, we will prove that for any $\varepsilon >0$ one can find $\delta >0$ such that $|\, C(\rho_{AB})^2 - 4 \det(\zeta_A)\, | < \varepsilon\,$ whenever $M\big(\rho_{AB}\big):=\inf_{\Psi_{AB}} D\big(|\Psi_{AB}\rangle \langle \Psi_{AB}|, \rho_{AB} \big) < \delta$. Here $D$ denotes the trace-distance defined for arbitrary states $\rho_1$ and $\rho_2$ via $D(\rho_1, \rho_2) = \frac{1}{2} \tr{|\rho_1-\rho_2|}$. 

The quantity $M\big(\rho_{AB}\big)$ is clearly a measure of ``mixedness'' of the state $\rho_{AB}$. To further enforce this terminology, it holds that 
\begin{eqnarray}
&&0\leq \frac{1}{4} \left( 1-\tr{\rho_{AB}^2} \right) \nonumber \\ 
&&= \frac{1}{4} \vert\, \tr{\big(|\Psi_{AB}\rangle \langle \Psi_{AB}|-\rho_{AB}\big) \big( |\Psi_{AB}\rangle \langle \Psi_{AB}| + \rho_{AB}\big)}\vert  \nonumber \\
&& \leq\frac{1}{2} \tr{\vert\, |\Psi_{AB}\rangle \langle \Psi_{AB}| - \rho_{AB}\,\vert} \nonumber \\
&&= D\big(|\Psi_{AB}\rangle \langle \Psi_{AB}|, \rho_{AB} \big)\,,
\end{eqnarray}
for any $|\Psi_{AB}\rangle$. Hence $\frac{1}{4} \left( 1-\tr{\rho_{AB}^2} \right) \leq M\big(\rho_{AB}\big)$, where $\tr{\rho_{AB}^2}$ is known as the purity of the state $\rho_{AB}$.

We begin by showing that concurrence is continuous with respect to the trace distance: in  proving this, we will closely follow the technique used in Ref.\,\cite{Guo2013}. Let us extend $C$ into the trace class of $\C^2 \otimes \C^2$ by defining
\begin{eqnarray}
\widetilde{C}(T) := \left\{ \begin{array}{ll}
\tr{| T|} C\big(\frac{|T|}{\tr{|T|}} \big),& T\neq 0 \\
0, & T=0  
\end{array}\right. \,.
\end{eqnarray}
For all $T$ the function $\widetilde{C}(T)$ can then be equivalently expressed as $\widetilde{C}(T) = \inf_{\{t_i, |\psi_i\rangle\}} \sum_i t_i C(\psi_i )$, where $t_i\geq 0$ satisfies $|T| = \sum_i t_i |\psi_i\rangle \langle \psi_i|$ and $||\psi_i||=1$ for all $i$. Assuming that $|T_1| \leq |T_2|$ it is then straightforward to see that
\begin{eqnarray}
\widetilde{C}(T_1) \leq \widetilde{C}(T_2)\,.
\end{eqnarray}
Let $\rho_1$ and $\rho_2$ be quantum states in $\C^2\otimes \C^2$ and define $\tau = \rho_1 - \rho_2$, so that $\rho_1 = | \tau  + \rho_2 | \leq  |\tau| + \rho_2 = |\,|\tau|  + \rho_2 |$. For any $\varepsilon>0$, one can find ensembles $\{t_i, |\psi_i\rangle\}$ and $\{p_j, |\varphi_j\rangle\}$ such that $|\tau| = \sum_i t_i |\psi_i\rangle \langle \psi_i|$ and $\rho_2 = \sum_j p_j |\varphi_j\rangle \langle \varphi_j|$ and satisfying $\widetilde{C}(|\tau|) \geq \sum_i t_i C(\psi_i) - \varepsilon/2$ and  $\widetilde{C}(\rho_2) \geq  \sum_j p_j C(\varphi_j) - \varepsilon/2$. Since $\sum_i \sum_j (t_i + p_j) = \tr{|\tau| + \rho_2|}$, we have
\begin{eqnarray}
C(\rho_1) &=& C(\tau + \rho_2) \leq \widetilde{C}(|\tau| + \rho_2) \nonumber \\
&\leq &  \sum_i t_i |\psi_i\rangle \langle \psi_i| + \sum_j p_j |\varphi_j\rangle \langle \varphi_j| \vert \nonumber \\
&\leq & \widetilde{C}(|\tau|) + C(\rho_2) + \varepsilon\,.
\end{eqnarray}
Because the relation holds for arbitrary $\varepsilon>0$, we can conclude that 
\begin{eqnarray}
|C(\rho_1) - C(\rho_2) | &\leq& \widetilde{C}(|\rho_1 - \rho_2|) \nonumber \\
&=& \tr{|\rho_1 - \rho_2|} C\left(\frac{|\rho_1 - \rho_2|}{\tr{|\rho_1 - \rho_2|}}\right) \nonumber \\
&\leq & 2 D(\rho_1, \rho_2)\,.
\end{eqnarray}

On the other hand, whenever $\rho_1$ and $\rho_2$ are states in $\C^2\otimes \C^2$, the reduced one-qubit states $\zeta_i = \text{tr}_{B}\left[\rho_i\right], i=1,2,$ satisfy
\begin{eqnarray}
\vert \det(\zeta_1) - \det(\zeta_2) \vert &=&\frac{1}{2} \vert \tr{\zeta_1^2}- \tr{\zeta_2^2} \vert \nonumber \\
&=& \frac{1}{2} \vert \tr{\big(\zeta_1-\zeta_2\big)\big(\zeta_1+\zeta_2\big)} \vert \nonumber \\
&\leq & 2 D\big(\zeta_1, \zeta_2\big) \leq  2 D\big(\rho_1, \rho_2\big)\,, \\
\nonumber
\end{eqnarray}
where we have used the property $2 \det(\zeta) = 1 - \tr{\zeta^2}$ that holds for all one-qubit states $\zeta$ and the data-processing inequality of trace-distance $D\big(\rho_1, \rho_2\big) \geq D\big(\mc E(\rho_1),\mc E(\rho_2) \big)$ that holds for all CPTP linear maps $\mc E$ (such as the partial trace).

Using the above relations, we can easily prove our claim. Let $\varepsilon >0$ and an unit vector $|\Psi_{AB}\rangle \in \C^2\otimes \C^2$ be arbitrary and denote $\rho_A = \text{tr}_{B}\left[|\Psi_{AB}\rangle \langle \Psi_{AB} | \right]$. We have
\begin{eqnarray}
&&\vert C(\rho_{AB})^2 - 4 \det(\zeta_A) \vert \nonumber \\
&& \leq \vert C(\rho_{AB})^2 - C(\Psi_{AB})^2 \vert +  \vert 4 \det(\rho_A) - 4 \det(\zeta_A) \vert \nonumber \\
&& \leq 2 \,\vert C(\rho_{AB}) - C(\Psi_{AB}) \vert + 4\,\vert  \det(\rho_A) -  \det(\zeta_A) \vert \nonumber \\
&& \leq 12 \,D\big(|\Psi_{AB}\rangle \langle \Psi_{AB}|, \rho_{AB}\big)\,,
\end{eqnarray}
and consequently $ \vert C(\rho_{AB})^2 - 4 \det(\zeta_A) \vert  \leq 12 \,M\big(\rho_{AB}\big)$ for all $\rho_{AB}$. Choosing $\delta = \frac{\varepsilon}{12}$ proves the claim. As a by-product we get the bounds
\begin{eqnarray}
4 C_-  \leq C(\rho_{AB})^2 \leq 4 C_+\,,
\end{eqnarray}
where $C_\pm := \det(\zeta_A) \pm 3 M\big(\rho_{AB}\big)$

In summary, we can conclude that $M\big(\rho_{AB}\big)\approx 0$ implies that $\rho_{AB}$ is both approximately pure ($\tr{\rho_{AB}^2}\approx 1$) and that $C(\rho_{AB})^2\approx 4 \det(\zeta_A)$.

\end{document}